\documentclass[]{spie}  

\usepackage{amsmath,amsfonts,amssymb}
\usepackage{graphicx}
\usepackage[version=4]{mhchem}
\usepackage{float}
\usepackage{bm}
\usepackage{upgreek}
\usepackage[colorlinks=true, allcolors=blue]{hyperref}

\hypersetup{breaklinks=true}   
\usepackage{xurl}              

\title{Characterization of a symmetric-facet dual-ruled grating for spatial heterodyne spectroscopy}

\author[a,\textdagger,*]{Cole Meyer}
\author[b,\textdagger]{Joanne Flores}
\author[a]{Jason Corliss}
\author[a]{Walter Harris}
\affil[a]{Lunar and Planetary Laboratory, University of Arizona, 1629 E University Blvd, Tucson, USA}
\affil[b]{Department of Electrical and Computer Engineering, University of Arizona, 1230 E Speedway Blvd, Tucson, USA}

\authorinfo{\textdagger\;These authors contributed equally to
this work.\\$^*$E-mail: cmmeyer@arizona.edu, Telephone: 1 320 296 9560}

\begin{document} 
\maketitle

\begin{abstract}
Dual-bandpass spatial heterodyne spectrometers (DB-SHS) enable simultaneous high-resolution measurements of widely separated passbands, providing powerful diagnostics of astrophysical and planetary environments. However, DB-SHS instruments require a single incident beam to span two adjacent diffraction gratings with distinct ruling densities and blaze angles, resulting in a large gap between ruled sections that reduces throughput. Dual-ruled gratings solve this problem by integrating multiple ruled panels onto a single substrate, minimizing the dead space between ruled sections. We present experimental validation of a first-generation symmetric-facet dual-ruled grating manufactured by Bach Research, mechanically ruled at $800$ and $\mathrm{2000\;gr\;mm^{-1}}$ with a $13.8^\circ$ blaze angle. Using a stabilized deuterium source alongside a Czerny-Turner monochromator, we measured diffraction efficiencies into the $m = 0, \pm1, \pm2$ orders from $200$ to $\mathrm{700\;nm}$. We compare these results with theoretical predictions from rigorous coupled-wave analysis (RCWA), inferring a facet asymmetry of $\lesssim1^\circ$ and $\sim70\%$ facet duty cycle indicative of minor manufacturing defects. This work demonstrates the viability of  mechanically ruled, symmetric-facet, dual-ruled gratings and lays the foundation for laboratory validation of the first DB-SHS, ultimately enabling high-resolution spectroscopy of distinct spectral regions relevant to astrophysical and planetary remote sensing.
\end{abstract}


\begin{figure}[t]
    \centering
    \begin{minipage}{0.56\textwidth}
        \centering
        \includegraphics[width=\linewidth]{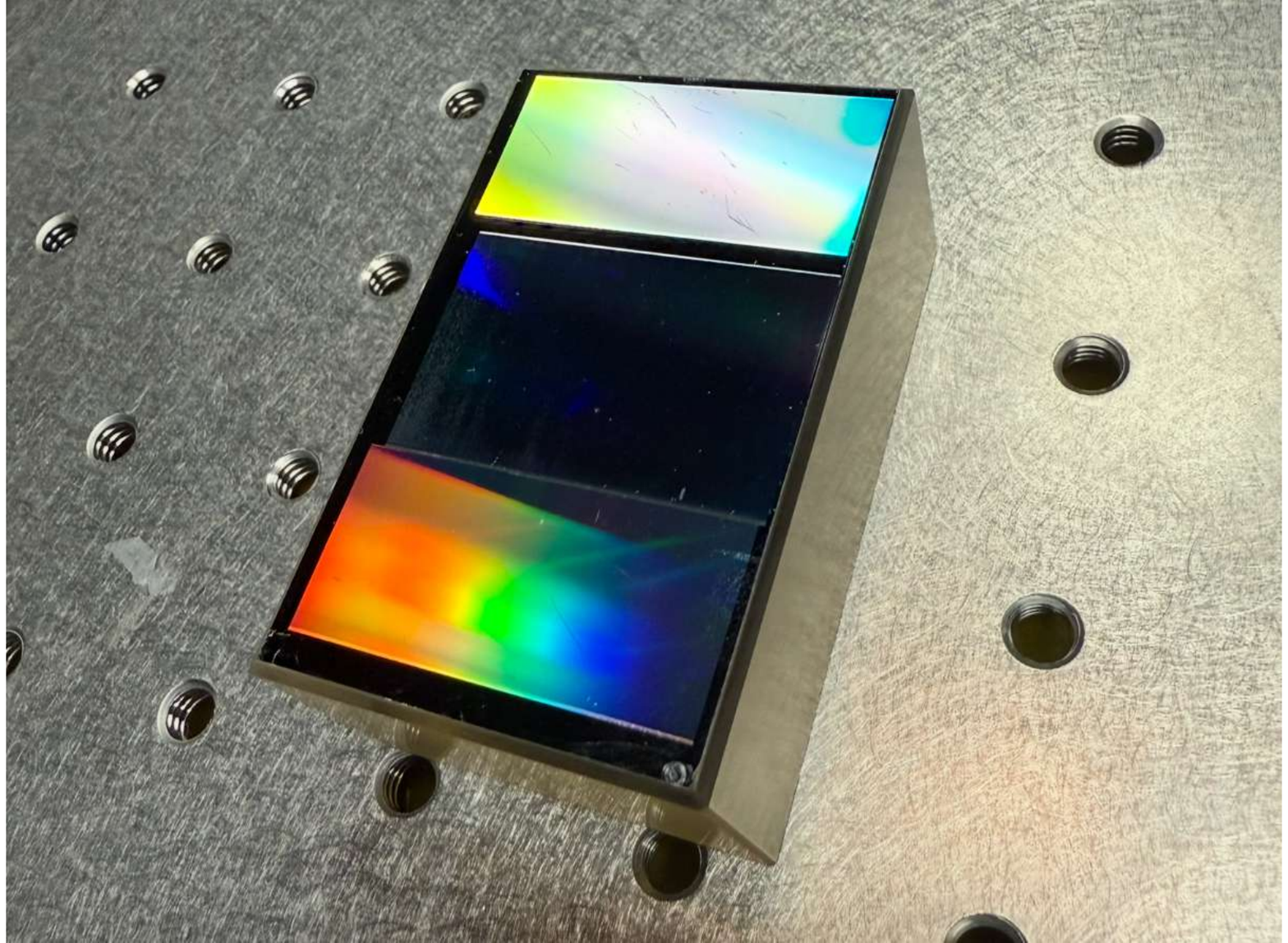}
    \end{minipage}\hfill
    \begin{minipage}{0.42\textwidth}
        \centering
        \includegraphics[width=\linewidth]{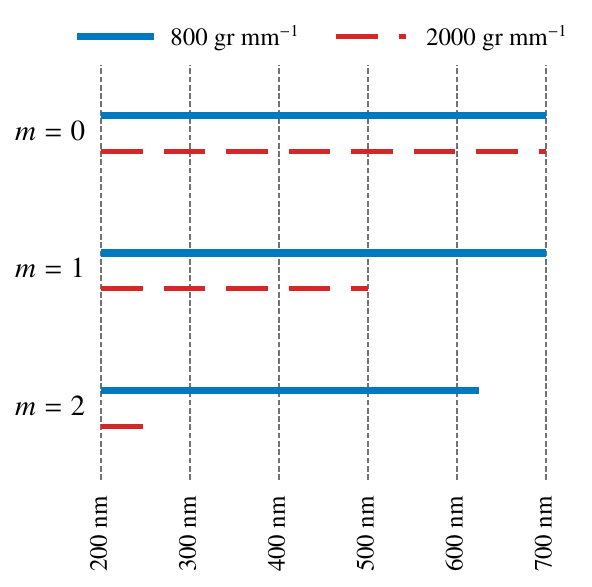}
    \end{minipage}
    
    \caption{Left: Photograph of the symmetric-facet dual-ruled grating characterized in this work. The top (panel 1) and bottom (panel 3) sections are ruled at $\mathrm{800\;gr\;mm^{-1}}$; the central section (panel 2) is ruled at $\mathrm{2000\;gr\;mm^{-1}}$. Right: Schematic indicating the wavelength ranges over which each diffraction order ($m = 0, 1, 2$) propagates for the $\mathrm{800\;gr\;mm^{-1}}$ (solid) and $\mathrm{2000\;gr\;mm^{-1}}$ (dashed) panels at normal incidence. Higher orders are evanescent beyond the cutoff wavelengths indicated by the bar endpoints.}
    \label{fig:diff_grating}
\end{figure}

\section{INTRODUCTION}~\label{sec:intro}

High-resolution spectroscopy is a fundamental tool for astrophysical and planetary remote sensing, enabling study of spectral features and line shapes across a broad range of physical environments. In numerous applications, physical parameters are derived from spectral regions that are widely separated in wavelength. For instance, the ratio of Ly$\alpha$ ($1215.7\mathrm{\;\AA}$) to Ly$\beta$ ($1025.7\mathrm{\;\AA}$) emission intensities distinguishes optically thin from optically thick gas~\cite{Hummer1992}. Conventionally, such observations rely on broadband low-resolution spectrometers, high-resolution scanning spectrometers, or echelle spectrographs. Broadband instruments sacrifice spectral resolution, while scanning Fourier transform spectrometers (FTS; e.g., scanning Michelson interferometers) introduce temporal offsets between observations of separated spectral regions, require precise re-registration to sample the same spatial area, and carry increased risk of mechanical failure and systematic error. Echelle spectrographs can observe separated regions at high resolution, but their small entrance apertures restrict them to point sources and demand large telescopes to compensate for reduced light-collecting efficiency. In most cases, mass and cost limitations disqualify the option of two separate instruments that may otherwise be capable of simultaneously observing multiple spectral regions at high resolution. A dual-bandpass spatial heterodyne spectrometer (DB-SHS), a variant of the well-studied spatial heterodyne spectrometer~\cite{HarlanderEA91_spatial,2015ApOpt..54.8835C}, offers an elegant solution to these challenges: a single instrument, consisting of no moving parts, capable of simultaneously obtaining high-resolution spectra over two widely separated passbands.
    
A DB-SHS can be implemented in two ways. In a \textit{multi-order} design, a single grating heterodynes several diffraction orders onto a shared detector, sampling multiple widely separated lines onto one interferogram. In a \textit{multi-ruling} design, the incident beam spans two distinct ruled regions with different groove densities and blaze angles, heterodyning each bandpass about its own heterodyne wavelength. The multi-ruling approach is generally preferable: it avoids the multiplexing disadvantage associated with a shared interferogram, in which shot noise from bright spectral features in one bandpass contributes to the noise floor of the other bandpass, degrading sensitivity to weak features. Moreover, in the multi-ruling design, each bandpass can be tuned independently toward targeted spectral regions rather than to the near-harmonic band centers imposed by a single-grating design. Realizing a multi-ruling design requires a single beam to span two adjacent ruled sections within a common optical path. Separate gratings could achieve this in principle, but their physical edges leave a large gap between ruled sections that fails to diffract incident light, significantly reducing throughput. A dual-ruled grating eliminates this gap by integrating both ruled sections onto a single substrate. Ideal behavior of an all-reflective SHS further benefits from symmetric grating facets, which ensures equal power into positive and negative orders. Because this monolithic, symmetric-facet ruling is a novel fabrication approach, its diffraction efficiency and facet quality must be verified before it can be incorporated into a DB-SHS. In this work, we characterize and validate the first symmetric-facet dual-ruled grating, manufactured by Bach Research specially for DB-SHS.

The structure of this paper is as follows. In Section~\ref{sec:methodology}, we describe our novel symmetric-facet dual-ruled grating and outline the experimental setup, data reduction, and theoretical modeling. In Section~\ref{sec:results}, we present the measured diffraction efficiencies and best-fit models. Additionally, we discuss implications of the measured efficiencies on the physical properties of the grating facets. Finally, we summarize our findings in Section~\ref{sec:conclusion}.

\section{METHODOLOGY}~\label{sec:methodology} 

\subsection{Dual-Ruled Grating Overview} 

Manufactured and mechanically ruled by Bach Research, the grating described here is dual-ruled and symmetrically blazed at $13.8^\circ$. As shown on the left in Figure~\ref{fig:diff_grating}, the substrate is divided into three equal panels: the outer two (1 and 3) are ruled at $\mathrm{800\;gr\;mm^{-1}}$, while the central panel (2) is ruled at $\mathrm{2000\;gr\;mm^{-1}}$. The grating surface carries a UV-enhanced Acton \#1200 coating: an aluminum layer $\geq\mathrm{70\;nm}$ thick protected by a $\mathrm{17\;nm}$ \ce{MgF2} overcoat. Applied to a flat mirror, this coating provides $\sim83-92\%$ reflectance over $200-700\mathrm{\;nm}$. We treat the Acton \#1200 grating coating as being composed entirely of Al since MgF2 is optically transparent at our wavelengths and not expected to affect diffraction efficiency.

Bach Research provided a preliminary report including diffraction efficiency measurements in the $\pm1$ orders at $\mathrm{405\;nm}$ for each panel. In the $+1$ and $-1$ orders, the reported efficiencies are $19\%$ and $24\%$ for panel 1, $14\%$ and $12.5\%$ for panel 2, and $19\%$ and $25\%$ for panel 3. Crucially, the grating's efficiency has not been measured at the ultraviolet wavelengths or higher orders ($m>1$) at which future DB-SHS instruments will operate, so these data neither fully characterize its diffraction efficiency nor permit assessment of potential fabrication-induced imperfections. To address these gaps, we measure the diffraction efficiency of all three panels across the $m = 0, \pm1, \pm2$ orders, spanning from $\mathrm{700\;nm}$ (the upper limit of our light source) down to the $\mathrm{200\;nm}$ air-transmission cutoff, and characterize how performance trends toward the ultraviolet.

\subsection{Experimental Setup}~\label{sec:setup}

\begin{figure}[t]
\centering
\includegraphics[width=1\textwidth]{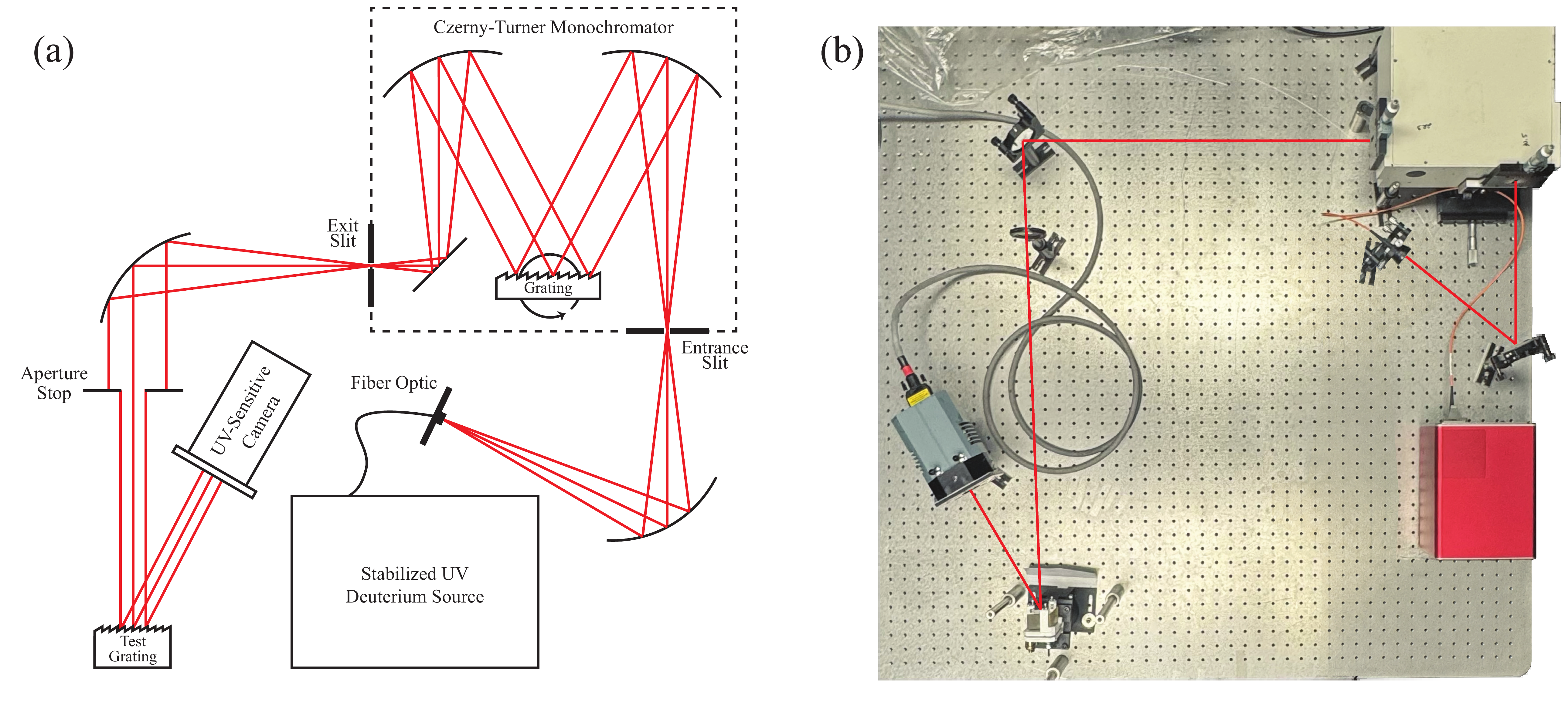}
\caption{(a) Annotated optical schematic of the diffraction efficiency measurement apparatus. Monochromatic light from a stabilized deuterium source is selected by a Czerny-Turner monochromator, collimated through an aperture stop, and directed either onto the test grating for diffracted-beam measurements or directly into the detector for incident-flux reference measurements. Both paths image the beam onto a UV-sensitive CCD camera. (b) Photograph of the laboratory implementation with the beam path highlighted in red.}\label{fig:experimental-setup}
\end{figure}

Diffraction-efficiency measurements were acquired at $\mathrm{200-700\;nm}$ in $\mathrm{100\;nm}$ increments. For each wavelength, two sets of images were collected: (1) the diffracted beams corresponding to the accessible diffraction orders ($m = 0, \pm1, \pm2$) for each grating panel, and (2) a reference measurement of the incident beam used for efficiency calculation (see Section~\ref{sec:diffeffcalc} for more details). Propagation of each diffraction order on each panel is depicted in Figure~\ref{fig:diff_grating} right.

A schematic and image of the experimental setup is shown in Figure~\ref{fig:experimental-setup}. We employed a stabilized deuterium light source with a broadband spectrum spanning $\mathrm{200-700\;nm}$. The deuterium source emission propagates into the entrance slit of a Czerny-Turner monochromator tuned to a desired wavelength before leaving through an exit slit. Using a bright Helium-Neon (HeNe) laser ($594.1\mathrm{nm}$), we estimate the wavelength range sampled by the exit slit to be $\Delta\lambda\sim12\mathrm{\;nm}$ based on laser line visibility across tuning. The quasi-monochromatic light then reflects off a concave folding mirror, collimating the beam and directing it toward an aperture stop. Finally, the beam propagates through one of two paths depending on which measurement type is desired. For diffracted beam measurements, the beam strikes the grating at normal incidence and diffracts into an Andor DV437 CCD detector with a \ce{MgF2} window. For reference beam measurements, the beam strikes the detector directly, providing a measurement of the incident flux.

\subsection{Data Reduction}

\subsubsection{Diffraction Efficiency Calculation}~\label{sec:diffeffcalc}

The diffraction efficiency $\eta_{m,\lambda}$ into order $m$ at wavelength $\lambda$ is defined as diffracted power $P_{\mathrm{diffr},m,\lambda}$ divided by incident power $P_{\mathrm{inc},m,\lambda}$:
\begin{align}\label{eqn:diffeff}
    \eta_{m,\lambda} &= \frac{P_{\mathrm{diffr},m,\lambda}}{P_{\mathrm{inc},m,\lambda}}.
\end{align}
We derive $P_{\mathrm{diffr},m,\lambda}$ and $P_{\mathrm{inc},m,\lambda}$ from two distinct experimental configurations: one in which the grating diffracts the monochromatic incident beam into the detector ($P_{\mathrm{diffr},m,\lambda}$) and one in which the beam strikes the detector directly ($P_{\mathrm{inc},m,\lambda}$). To calculate $P_{\mathrm{diffr},m,\lambda}$ and $P_{\mathrm{inc},m,\lambda}$ from beam images, we apply \texttt{median\_filter} from the \texttt{Python} package \texttt{scipy} to remove cosmic rays, hot pixels, and dead pixels, subtract image backgrounds (see Section~\ref{sec:straylight}), sum up all counts, and divide by exposure time.

We report relative diffraction efficiencies and do not attempt to isolate the \textit{absolute} diffraction efficiency (that of the uncoated grating) by removing the coating reflectance. In principle, one could cancel the coating contribution by referencing the incident beam against a pickoff mirror bearing the same Acton \#1200 coating. In practice this fails: the coating on a grating fills the grooves, yielding a non-flat surface whose thickness varies across each facet, and so cannot be matched by a flat mirror coating. Because the two coatings do not cancel, the literature instead typically reports the absolute efficiency and coating reflectance together as a single \textit{relative} diffraction efficiency, which is also the more useful quantity for instrument design. Hereafter we refer to these relative efficiencies simply as ``diffraction efficiencies.''

\subsubsection{Stray Light Subtraction}\label{sec:straylight}

To derive the count rate of each beam, as described in Section~\ref{sec:diffeffcalc}, it is crucial to effectively subtract the background of each beam image. Each exposure contains (1) the incident beam, (2) a bias offset, (3) dark counts, and (4) stray light scattered off optical components and diffuse emission from the light source. Despite extensive efforts, we were unable to completely eliminate stray light from our exposures given the particulars of our experimental setup. Moreover, we found that the stray light component contributed significantly, often exceeding the dark counts. We explored two different methods for stray light mitigation.

As a first attempt, we collected a ``dark'' frame for each $(m,\lambda)$ configuration in which the detector was left uncovered and the beam was translated $\sim\mathrm{4\;cm}$ off the detector chip. In this case, the beam would not be detected but much of the stray light that was captured in the beam image would still be captured by the dark frame. While typically effective for incident beam images, the approach proved ineffective and time intensive for diffracted beam images, especially at ultraviolet wavelengths where alignment of each $(m,\lambda)$ configuration took considerable time to complete.

\begin{figure}[t]
\centering
\includegraphics[width=1\textwidth]{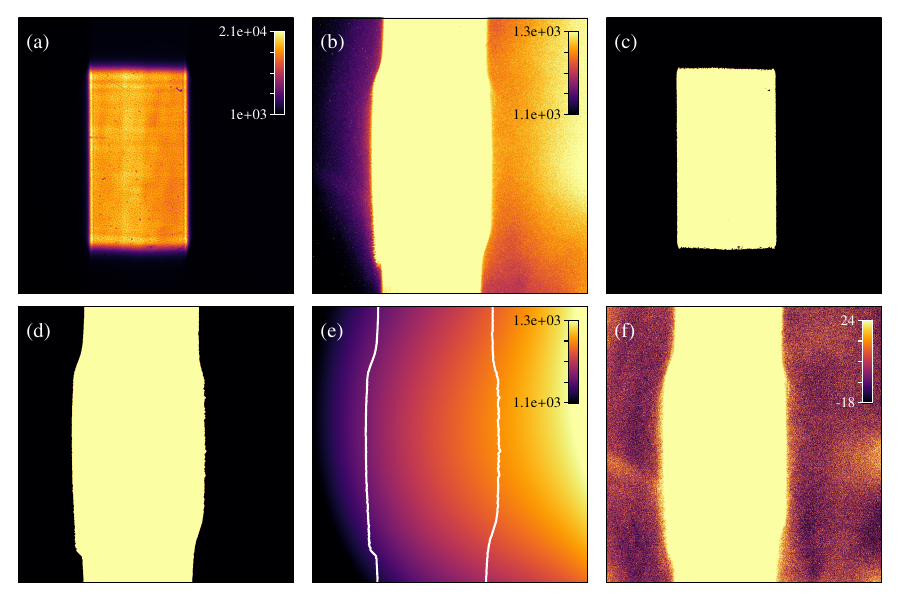}
\caption{Iterative stray light subtraction (Section~\ref{sec:straylight}) for a representative beam image. (a) Original CCD frame. (b) Same frame at reduced stretch, revealing the diffuse stray light pedestal. (c) Binary mask from the high (Otsu) threshold, isolating the bright beam core. (d) Expanded mask from hysteresis thresholding, grown outward from the high-threshold seed to capture faint beam wings. (e) Best-fit 2D polynomial background over the unmasked region; beam mask boundary in white. (f) Background-subtracted image at low stretch, showing near-zero residuals outside the beam footprint. In this example, the fitted background over the footprint is $\sim12\%$ of total counts, and non-beam residuals have standard deviation $\sigma_{bg}\approx21$
 counts ($\sim0.2\%$ of mean beam intensity), indicating effective background subtraction.}\label{fig:bg-sub}
\end{figure}

To remedy these challenges, we also explored computational background removal (see Figure~\ref{fig:bg-sub} for a visual depiction of the described procedure). We employed an iterative background removal process using hysteresis thresholding followed by 2D polynomial fitting of the image background. In each iteration, we applied the following procedure to the working image (initialized to the original beam image and updated with each pass):
\begin{enumerate}
    \item Convolve the working image with a Gaussian kernel using \texttt{gaussian\_filter} from \texttt{scipy} to remove noise spikes that may otherwise affect thresholding.
    \item Calculate two thresholds: (1) a high threshold via Otsu's method~\cite{otsu1975threshold} applied to the smoothed image using \texttt{threshold\_otsu} from \texttt{skimage}, targeting the bright beam core, and (2) a tunable low threshold $\tilde{x}+k_\mathrm{low}\sigma_x$, where $\tilde{x}$ and $\sigma_x$ are the background counts' median and standard deviation, respectively, and $k_\mathrm{low}$ is a tunable parameter typically $\sim1-5$.
    \item Beam pixels are identified via hysteresis: connected components in the low-threshold mask are retained only if they contain at least one pixel exceeding the high threshold, effectively seeding on the bright core and spreading outward into the fainter beam wings.
    \item Fit the un-masked background using the least squares method. Subtract the background from the working image.
\end{enumerate}
This approach proved highly effective across all $(m,\lambda)$ configurations, and thus, we used it in place of typical dark subtraction for all results presented in Section~\ref{sec:results}.

\begin{figure}[t]
\centering
\includegraphics[width=1\textwidth]{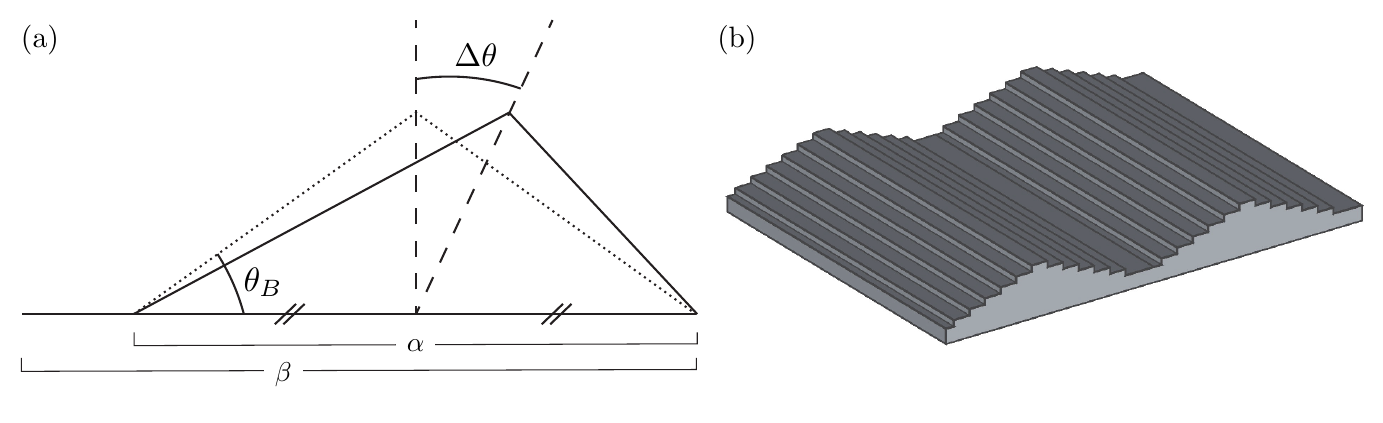}
\caption{Facet geometry used in RCWA forward modeling. (a) Cross-sectional schematic of a single grating period defining the blaze angle $\theta_B$, the facet asymmetry parameter $\Delta\theta$ (the difference in angle between the two facet faces relative to the ideal symmetric profile, indicated by dotted lines), and facet duty cycle $D=\alpha/\beta$. (b) Three-dimensional rendering of the discrete-strata approximation of two grating periods as implemented in \texttt{GD-CALC}, shown here with $N_\mathrm{strata}=7$ for visual clarity; all production models use $N_\mathrm{strata}=50$.}\label{fig:facet-schematic}
\end{figure}

\subsubsection{Error Estimation}~\label{sec:error}

In Section~\ref{sec:results}, we report uncertainties on all measured diffraction efficiencies. These uncertainties include two components calculated from the beam images: (1) absolute photon noise $\sigma_p = \sqrt{N}$, where $N$ is the total integrated counts over the beam footprint, and (2) absolute background subtraction uncertainty $\sigma_{bg}\sqrt{M}$, where $\sigma_{bg}$ is the pixel-to-pixel standard deviation of the non-beam region of the beam-subtracted image and $M$ is the number of pixels in the beam footprint. These sources are converted to the rate domain by dividing by the exposure time $t$ and added in quadrature to find the total absolute rate uncertainty:
\begin{align*}
    \sigma_{P}^2 = (\sigma_p/t)^2 + (\sigma_{bg}\sqrt{M}/t)^2.
\end{align*}
Based on lab measurements, we estimate and incorporate an absolute ``systematic'' error $\sigma_{\eta,sys}=5\%$ on the diffraction efficiency designed to capture experimental variability. Systematic error $\sigma_{\eta,sys}$ includes, for example, time-varying source intensity, varying angles of incidence on the detector for different experimental configurations, and beam clipping by the edge of the detector chip. These three sources of uncertainty are then propagated into the corresponding diffraction efficiencies using standard Gaussian error propagation:
\begin{align*}
    \sigma_\eta^2 = \eta^2\left[\left(\frac{\sigma_{P,\mathrm{refl}}}{P_\text{refl}}\right)^2 + \left(\frac{\sigma_{P,\mathrm{diff}}}{P_\text{diff}}\right)^2\right] + \sigma_{\eta,sys}^2
\end{align*}
In practice, $\sigma_p$ and $\sigma_{bg}$ are typically negligible compared to $\sigma_{\eta,sys}$ so that $\sigma_\eta\sim5\%$.

\subsection{Theoretical Modeling}

To physically interpret the measured diffraction efficiencies in the context of potential manufacturing defects, we apply a generalized version of rigorous coupled-wave analysis~\cite{Moharam1981_rigorous} (RCWA) to symmetrically-blazed gratings. The \texttt{MATLAB}-based toolkit \texttt{GD-CALC}~\cite{Johnson2026_grating} (Grating Diffraction Calculator) offers a convenient RCWA implementation and has been applied extensively to infer diffraction grating properties such as pitch and duty cycle~\cite{Baker2025_fabrication, Belousov2019_laser,Belousov2020_determination}. Within \texttt{GD-CALC}, gratings are constructed using discrete layers, or \textit{strata}, over one period of the facet profile. Higher $N_\mathrm{strata}$ values improve accuracy while increasing computational time. RCWA represents the periodic structure and electromagnetic field using Fourier series, which can be truncated to improve computational time at the expense of reduced accuracy. We represent each triangular facet using $N_\mathrm{strata}=50$ with varying widths to form facet profiles (see Figure~\ref{fig:facet-schematic}(b), which uses $N_\mathrm{strata}=7$ for visualization purposes); larger $N_\mathrm{strata}$ values produce only negligible changes to predicted diffraction efficiencies. Similarly, we find that $2m_\mathrm{max}+1=11$ retained diffraction orders ($m=-5,-4,...,+4,+5$) are sufficient to converge diffraction efficiencies for the orders of interest ($m=-2,-1,0,+1,+2$). We restrict our modeling efforts to normal incidence, consistent with the experimental setup. As previously mentioned, we treat the Acton \#1200 coasting as purely aluminum for modeling purposes. Optical constants (extinction coefficient and index of refraction) for Al are drawn from Ref.~\citenum{Rakic1995_algorithm}. Emission from the broadband deuterium source is unpolarized and thus, we report diffraction efficiencies for unpolarized light.

\begin{figure}[t]
\centering
\includegraphics[width=1\textwidth]{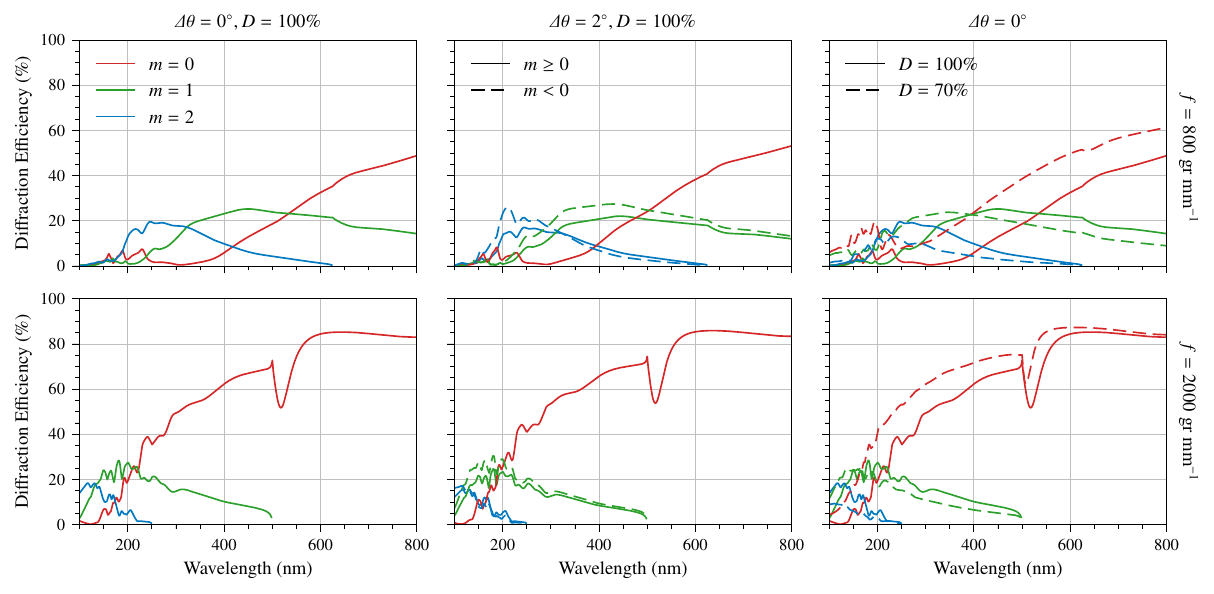}
\caption{RCWA model predictions illustrating the separable effects of facet asymmetry $\Delta\theta$ and facet duty cycle $D$ on diffraction efficiency for $f=\mathrm{800\;gr\;mm^{-1}}$ (top) and  $f=\mathrm{2000\;gr\;mm^{-1}}$ (bottom) at normal incidence. Left: Baseline symmetric, full duty cycle case ($\Delta\theta=0^\circ, D=100\%$), where $m>0$ and $m<0$ orders are degenerate. Center: Effect of introducing facet asymmetry ($\Delta\theta=2^\circ,D=100\%$), which breaks the $\pm m$ degeneracy and redistributes power between positive and negative orders. Right: Effect of introducing $D<100\%$ ($\Delta\theta=0^\circ; D=100\%\mathrm{\;vs.\;}70\%$), which suppresses efficiency in the blaze orders and redistributes power into $m=0$ and higher orders. The distinct signatures of $\Delta\theta$ and $D$ motivate their joint retrieval via MCMC.}\label{fig:model-demo}
\end{figure}

Construction of facet profiles requires four input parameters: (1) groove density $f$, (2) design blaze angle $\theta_B$, (3) facet asymmetry $\Delta\theta$ (see Figure~\ref{fig:facet-schematic}(a) for our $\Delta\theta$ definition), and (4) facet duty cycle $D$, where we define facet duty cycle as the ratio of facet width to grating pitch (see Figure~\ref{fig:facet-schematic}(a)). In the case of our symmetric-facet dual-ruled grating, we have $f=\mathrm{800\;gr\;mm^{-1}}$ and $f=\mathrm{2000\;gr\;mm^{-1}}$ across panels, and $\theta_B=13.8^\circ$ for all panels. Importantly, our model of the facet structure is merely an approximation of what are likely to be complex and spatially variable manufacturing defects, and therefore, our model is used only to capture large-scale, spatially-averaged behavior. We discuss the implications of the fitted parameters for the physical groove structure alongside their caveats in Section~\ref{sec:results-params}.

Figure~\ref{fig:model-demo} illustrates that facet asymmetry $\Delta\theta$ and facet duty cycle $D$ imprint distinct, separable signatures on the diffraction efficiency, motivating the retrieval approach described below.

To retrieve $\Delta\theta$ and $D$ from measured diffraction efficiencies, we generate a grid of RCWA models spanning $\Delta\theta\in[-5,5^\circ]$ and $D\in[0,100\%]$ for both $f=\mathrm{800\;gr\;mm^{-1}}$ and $f=\mathrm{2000\;gr\;mm^{-1}}$. Because direct RCWA evaluation is computationally expensive, we construct a surrogate model by interpolating over this grid using \texttt{RegularGridInterpolator} from \texttt{scipy}, enabling rapid forward model evaluation for arbitrary ($\Delta\theta,D$) inputs within the grid boundaries. We then perform parameter retrieval via Markov Chain Monte Carlo (MCMC) using the \texttt{Python} package \texttt{emcee}. At each MCMC step, the forward model queries the surrogate for theoretical diffraction efficiencies, applies boxcar smoothing to the result to match the monochromator exit slit width, and evaluates the likelihood. We find that 12 walkers (6 per free parameter), 2000 burn-in steps, and 5000 production steps yield sufficient convergence. Retrieved values are reported in Section~\ref{sec:results}, with corner and trace plots provided in Appendix~\ref{appendix:mcmc}.

The full reduction and fitting pipeline is publicly available (see Code and Data Availability).

\newpage
\section{RESULTS \& DISCUSSION}~\label{sec:results}

\begin{figure}[t]
\centering
\includegraphics[width=0.8\textwidth]{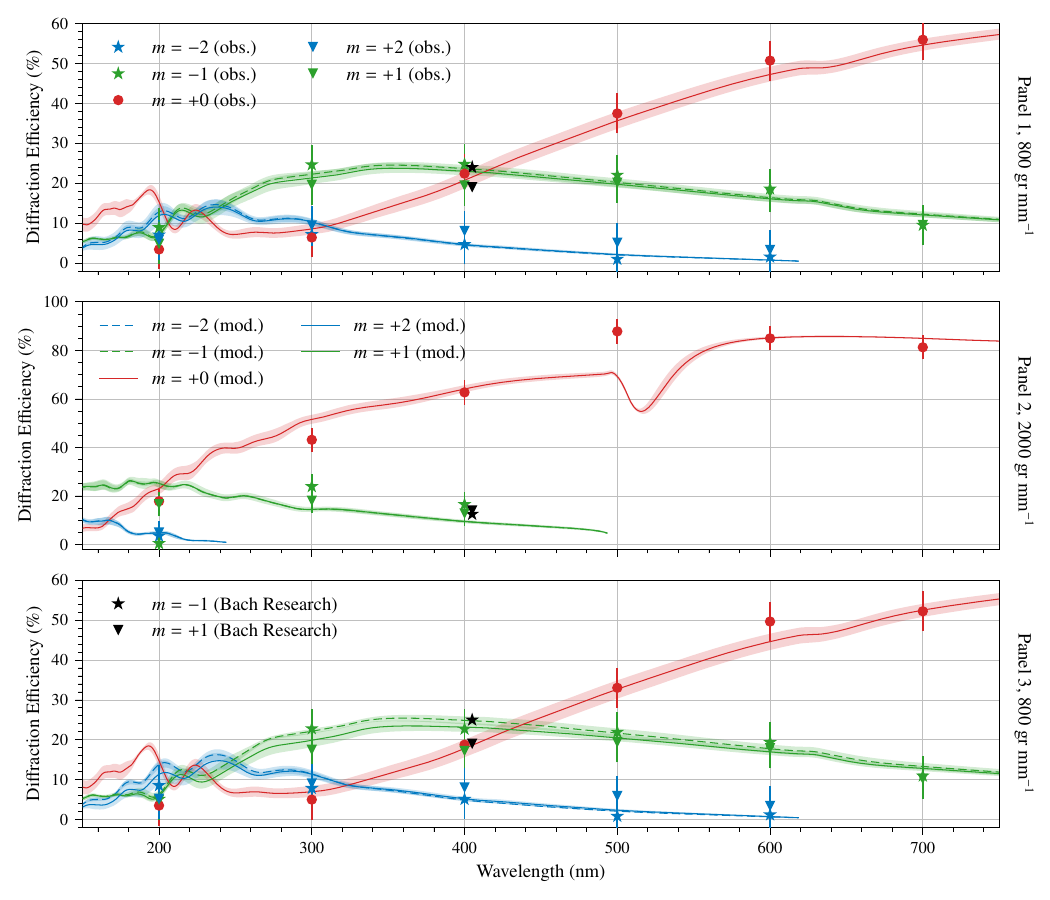}
\caption{Measured (points with error bars) and best-fit MCMC model (lines) diffraction efficiencies for orders $m=-2,-1,0,+1,+2$ at $\mathrm{200-700\;nm}$, shown for each of the three grating panels. Top: Panel 1, $\mathrm{800\;gr\;mm^{-1}}$. Middle: Panel 2, $\mathrm{2000\;gr\;mm^{-1}}$. Bottom: Panel 3, $\mathrm{800\;gr\;mm^{-1}}$. Error bars represent the combined photon noise, background subtraction uncertainty, and 5\% systematic floor described in Section~\ref{sec:error}. Measured efficiencies reported by the Bach Research report are shown in all three panels for $m=\pm1$ in black. Best-fit parameters ($\Delta\theta,D$) are given in Section~\ref{sec:results}, and the full MCMC diagnostics are provided in Appendix~\ref{appendix:mcmc}.}\label{fig:mcmc-keystone}
\end{figure}

\subsection{Diffraction Efficiency}

We report diffraction efficiencies in $m=0,\pm1,\pm2$ across all three panels (including $f=\mathrm{800\;gr\;mm^{-1}}$ and $f=\mathrm{2000\;gr\;mm^{-1}}$) at $\mathrm{200-700\;nm}$ in Figure~\ref{fig:mcmc-keystone} alongside the best fit models for each panel.

\subsubsection{Panels 1 \& 3 (\boldmath$\mathrm{800\;gr\;mm^{-1}}$)}

The measured diffraction efficiencies in panels 1 and 3 are $\sim3-56\%$ for $m=0$, $\sim5-25\%$ for $m=\pm1$ (peaking at $\mathrm{300\;nm}$), and $\sim1-9\%$ for $m=\pm2$ (peaking at $\mathrm{300\;nm}$) with the summed diffraction efficiency $\eta_{tot}'=\eta_0 + \eta_{\pm1} + \eta_{\pm2}$ of panel 1 exceeding that of panel 3 by $\sim6\%$. There appears to be a moderate asymmetry between $m>0$ and $m<0$ based purely on the measured diffraction efficiencies; averaged across wavelengths and the two panels, $\eta_{+1} = 0.85\,\eta_{-1}$ and $\eta_{-2} = 0.69\,\eta_{+2}$. In general, the measured efficiencies were fit reasonably well with reduced chi-squared $\chi^2_{red,0}\sim1.0$, $\chi^2_{red,\pm1}\sim0.2$, and $\chi^2_{red,\pm2}\sim0.4$, averaged over panels 1 and 3. Our measured efficiencies for panels 1 and 3 agree well with those reported by Bach Research at 405 nm.

\subsubsection{Panel 2  (\boldmath$\mathrm{2000\;gr\;mm^{-1}}$)}

The diffraction efficiencies for panel 2 are $\sim18-81\%$ for $m=0$, $\sim1-24\%$ for $m=\pm1$, and $\sim4-5\%$ for $m=\pm2$. It is challenging to deduce to what extent an asymmetry might be present, if any, given what appears to be a spurious $m=-1$ measurement at 200 nm. Based on the other available measurements, there may be a minor asymmetry present, with $\eta_{+1} = 0.76\,\eta_{-1}$ and $\eta_{-2} = 0.72\,\eta_{+2}$ averaged over wavelength. Although the model captured the large-scale behavior of the measured efficiencies, the fits struggled compared to panels 1 and 3 with $\chi^2_{red,0}\sim3.3$, $\chi^2_{red,\pm1}\sim5.6$, and $\chi^2_{red,\pm2}\sim0.1$, where $\chi^2_{red,\pm2}$ is artificially low due to the presence of only one measurement in $m=2$. Note that the pitch of the $\mathrm{2000\;gr\;mm^{-1}}$ section of the grating, $\mathrm{500\;nm}$, is on the order of our wavelengths, and thus, we expect the presence of some cavity-like resonances that will impact the measured diffraction efficiencies. With the available information, we are unable to deduce whether these resonances are responsible for the stronger mismatch between model and measurement in panel 2 compared to panels 1 and 3. Our measured efficiencies agree well with those reported by Bach Research at 405 nm.

\subsection{Facet Asymmetry and Duty Cycle}~\label{sec:results-params}

For all three panels, we achieve tight, well-converged Markov Chain Monte Carlo (MCMC) chains with a typical acceptance fraction of $\sim70\%$ and autocorrelation times of $36$, $34$, and $32$ steps for panels 1-3, respectively, corresponding to an effective sample size of $\sim1700-1900$ (see Appendix~\ref{appendix:mcmc} for corner and chain plots for each panel). For panels 1-3, we find best-fitting physical parameters of $\Delta\theta = -0.4_{-1.0}^{+0.9}, 0.1_{-1.1}^{+1.1}, -0.9_{-1.0}^{+1.0}$ degrees and $D = 72.0_{-2.2}^{+2.2}, 92.3_{-6.3}^{+5.0}, 73.5_{-2.2}^{+2.3}\;\%$, respectively.

The inferred facet asymmetry ($\Delta\theta \neq 0$) and duty cycle ($D\neq100\%$) are consistent with the sequential mechanical ruling process used to manufacture the grating. To achieve a symmetrically blazed profile, the ruling engine first shaped one face of the groove using  a symmetrically triangular  diamond tip standard for conventional ``sawtooth'' blazed gratings, before indexing to rule the opposing face. During this second pass, the delicate apex formed by the initial ruling is highly susceptible to mechanical tearing, tool-drag, and plastic deformation of the gold master substrate. This secondary stress could structurally alter the groove profile, producing both the slight angular asymmetry and the prominent, flattened features modeled with $D\neq100\%$. Any deviation from normal incidence in the tool setup could also propagate an asymmetry. These deformations are likely to be complex and spatially non-uniform across the substrate, preventing them from being fully captured by our idealized RCWA model. Still, extracting these macroscopic effective parameters remains crucial for accurately predicting the optical throughput and overall performance of a future DB-SHS instrument.

\subsection{Implications for DB-SHS}

The measured efficiencies and inferred facet properties carry several implications for a future multi-ruling DB-SHS. Facet asymmetry most directly reduces fringe contrast, and hence signal-to-noise ratio (S/N): an all-reflective SHS interferes the $+m$ and $-m$ diffracted beams, so the modulated (fringe-carrying) signal scales as $\sqrt{\eta_{+m}\eta_{-m}}$ while the total flux scales as $\eta_{+m}+\eta_{-m}$ (see Ref.~\citenum{HarlanderEA91_spatial}). Only the matched, lower-efficiency fraction contributes to the interference; the excess power from the brighter order forms an un-modulated background that the transform removes but that still raises the shot-noise floor across all channels contributing to the multiplexing disadvantage. This penalty is small for the asymmetries we infer. For $\eta_{+1}=0.85\eta_{-1}$ on the $800\mathrm{\;gr\;mm^{-1}}$ panels, the fringe contrast $V=2\sqrt{\eta_{+1}\eta_{-1}}/(\eta_{+1}+\eta_{-1})$ drops below unity by only $\sim0.3\%$. Moreover, the efficiency asymmetry can be compensated in SHS by rotating the pilot optics given the small $\Delta\theta\sim1^\circ$ retrieved. One subtlety is that the $800\mathrm{\;gr\;mm^{-1}}$ band is served by two physically distinct rulings (panels 1 and 3) feeding opposite arms, so any difference in their facet profiles adds to this intrinsic $\pm m$ imbalance, though this effect is difficult to quantify for SHS without measurements of off-normal efficiencies. Panels 1 and 3 nonetheless return consistent parameters, and the single central panel's near-symmetric profile implies well-balanced arms for the $2000\mathrm{\;gr\;mm^{-1}}$ band.

A second consideration is that because a DB-SHS records both passbands simultaneously in a single integration, the exposure time is set by the least efficient interaction among the spanned panels and orders (saturation aside). For instance, operating the central $2000\mathrm{\;gr\;mm^{-1}}$ panel in $m=\pm2$ at $200\mathrm{\;nm}$, where we measure $\sim4-5\%$ efficiency, would demand a long integration to reach adequate S/N regardless of whether panels 1 and 3 were more efficient in the companion band. Therefore, while separating the passbands into distinct interferograms reduces the multiplexing disadvantage for imbalanced passband intensities compared to a multi-order DB-SHS, the challenge of balancing exposure time between passbands persists for both DB-SHS implementations.

Finally, the non-ideal facet structure inferred above will scatter some fraction of the incident light out of the intended diffraction orders, producing stray light within the instrument. We do not expect this component to significantly degrade DB-SHS performance for two reasons. First, scattered light is distributed over a broad range of angles so that only a small fraction is redirected toward the detector and falls within the acceptance angle of the downstream exit optics; the remainder is rejected by stops and baffling. Second, an SHS is intrinsically effective at suppressing out-of-band and incoherent light: only light near the heterodyne wavelength produces resolved, low-frequency fringes, whereas diffuse scattered light contributes primarily to the un-modulated background and is therefore removed in the transform while adding only weakly to the noise floor. A precise estimate of the stray-light level reaching the detector requires end-to-end optical modeling of the full DB-SHS, but the combination of varied scattering angles and SHS out-of-band self-filtering suggests that scattering from the measured facet imperfections will be only a minor effect.

\section{CONCLUSION}~\label{sec:conclusion}

We have presented experimental characterization and validation of the first symmetric-facet dual-ruled grating, mechanically ruled by Bach Research at $800$ and $2000\mathrm{\;gr\;mm^{-1}}$ for use in a future dual-bandpass spatial heterodyne spectrometer (DB-SHS). Using a stabilized deuterium source and a Czerny-Turner monochromator, we measured the diffraction efficiency of all three ruled panels into the $m = 0, \pm1, \pm2$ orders from $200$ to $700\mathrm{\;nm}$, applied computational stray-light subtraction based on hysteresis thresholding and 2D polynomial background fitting, and interpreted the results with RCWA forward models, retrieving the effective facet asymmetry $\Delta\theta$ and duty cycle via MCMC. Our main findings are:
\begin{itemize}
    \item The $800\mathrm{\;gr\;mm^{-1}}$ panels reach efficiencies of $\sim5-25\%$ in $m = \pm1$ and $\sim1-9\%$ in $m = \pm2$ (both peaking near $300\mathrm{\;nm}$), with up to $\sim56\%$ in $m = 0$, while the $2000\mathrm{\;gr\;mm^{-1}}$ panel reaches $\sim1-24\%$ in $m = \pm1$, $\sim4-5\%$ in $m = \pm2$, and up to $\sim81\%$ in $m = 0$; efficiencies generally decline toward the ultraviolet.
    \item A modest $\pm m$ asymmetry is present in all panels: $\eta_{+1} \approx 0.85\,\eta_{-1}$ and $\eta_{-2} \approx 0.69\,\eta_{+2}$ for the $800\mathrm{\;gr\;mm^{-1}}$ panels, and $\eta_{+1} \approx 0.76\,\eta_{-1}$ and $\eta_{-2} \approx 0.72\,\eta_{+2}$ for the $2000\mathrm{\;gr\;mm^{-1}}$ panel.
    \item The fits favor small facet asymmetries ($|\Delta\theta| \lesssim 1^\circ$) and reduced duty cycles ($\sim73\%$) on the outer $800\mathrm{\;gr\;mm^{-1}}$ panels; the two outer panels return consistent parameters, as expected for their shared nominal specification and symmetric placement. The middle panel is fitted best by negligible asymmetry and high duty cycle ($\sim92\%$).
    \item The inferred asymmetries and duty cycles are consistent with the sequential mechanical ruling of a symmetric blaze, in which the second ruling pass damages the apex formed by the first, plausibly producing both effects simultaneously.
    \item The measured asymmetries impose only a small fringe contrast penalty ($\sim0.3\%$) for DB-SHS that can be further compensated in the pilot optics; DB-SHS throughput will instead be limited by the least efficient interaction among the spanned panels and orders, and scattering from the inferred imperfections is expected to be a minor contaminant.
\end{itemize}
These results demonstrate the viability of monolithic symmetric-facet dual-ruled gratings and establish a validated performance baseline for the principal optical component of a DB-SHS. The retrieved facet parameters capture the large-scale, spatially-averaged behavior of the grating well, even if they cannot fully describe its complex manufacturing defects. With the grating now characterized, the path forward is to integrate it into a prototype instrument and pursue the first laboratory validation of a DB-SHS.

\section*{CODE AND DATA AVAILABILITY}~\label{sec:code}

The analysis pipeline used to produce the results in this paper is publicly available on GitHub at \url{https://github.com/colemeyer/dr-grat-char} and is archived, together with the supporting data (raw detector frames and pre-computed model grids), at Zenodo (\href{https://doi.org/10.5281/zenodo.20682543}{\nolinkurl{doi:10.5281/zenodo.20682543}}). The diffraction-efficiency forward model was computed with GD-CALC,~\cite{Johnson2026_grating} available separately at \href{https://doi.org/10.24433/CO.7479617.v3}{\nolinkurl{doi:10.24433/CO.7479617.v3}}. This work employs \texttt{Python} packages \texttt{astropy}~\cite{2022ApJ...935..167A}, \texttt{emcee}~\cite{emcee_citation}, \texttt{funkyfresh}~\cite{milby_funkyfresh_2022}, \texttt{matplotlib}~\cite{Hunter:2007}, \texttt{numpy}~\cite{harris2020array}, \texttt{scipy}~\cite{2020SciPy-NMeth}, and \texttt{skimage}~\cite{van2014scikit}.

\acknowledgments 

We acknowledge support from the National Science Foundation (NSF) through a Graduate Research Fellowship under grant No. DGE-2137419 to C.M. The material contained in this document is based upon work supported by a National Aeronautics and Space Administration (NASA) cooperative agreement 80NSSC25M7084 to C.M. and also under grants No. 80NSSC23K15863054580 (PICASSO) and 80NSSC24K02990 (APRA) to J.C. Any opinions, findings, conclusions or recommendations expressed in this material are those of the authors and do not necessarily reflect the views of NASA or NSF. The authors used Claude Opus 4.8 for language and grammar clean-up and code readability, in accordance with SPIE and Committee on Publication Ethics (COPE) guidelines. We thank Mike Eiklenborg and Javier Alday for supporting our experimental setup.

\bibliography{refs} 
\bibliographystyle{spiebib} 

\appendix
\section{MARKOV CHAIN MONTE CARLO (MCMC) OUTPUTS} 
\label{appendix:mcmc}

\begin{figure}[H]
\centering
\includegraphics[width=0.9\textwidth]{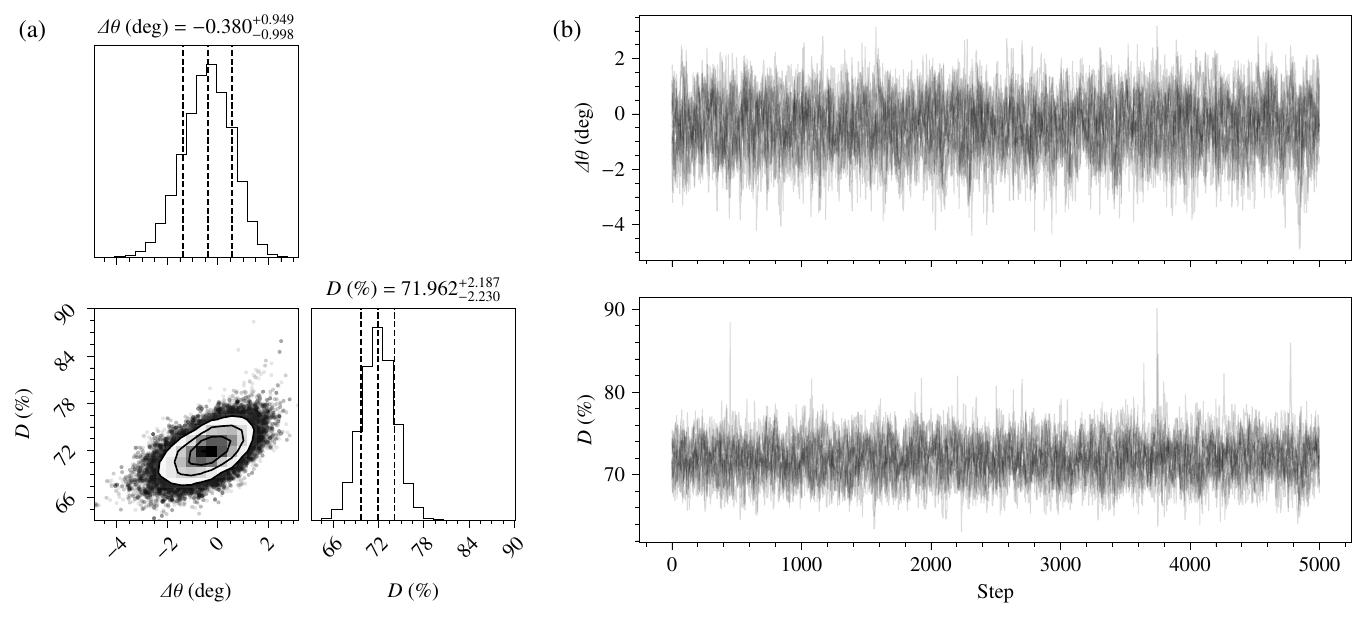}
\caption{MCMC posterior diagnostics for panel 1 ($\mathrm{800\;gr\;mm^{-1}}$). (a) Corner plot showing the 1D marginalized posteriors and 2D joint posterior for $\Delta\theta$ and $D$; dashed lines indicate the 16th, 50th, and 84th percentiles. The well-localized, approximately Gaussian joint posterior indicates good parameter constraint with weak covariance between $\Delta\theta$ and $D$. (b) Trace plots for all 12 walkers across 5000 production steps, confirming chain convergence and good mixing.}\label{fig:mcmc-app-panel1}
\end{figure}

\begin{figure}[H]
\centering
\includegraphics[width=0.9\textwidth]{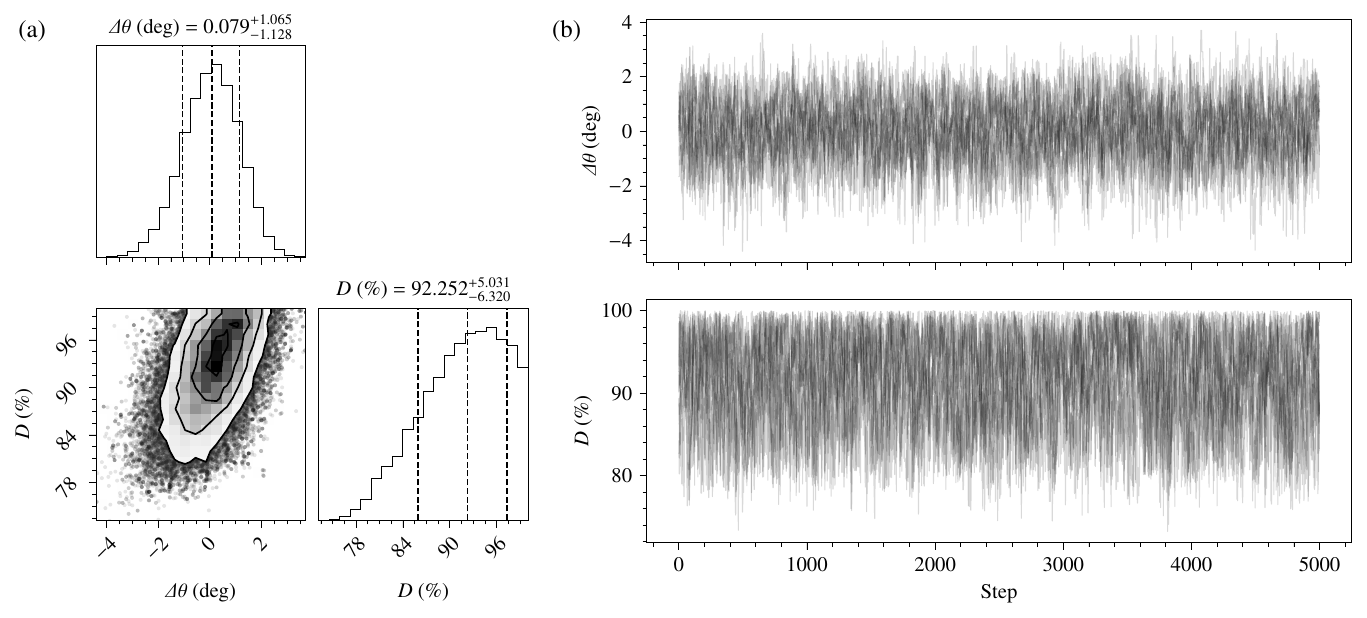}
\caption{Same as Figure~\ref{fig:mcmc-app-panel1} but for panel 2 ($\mathrm{2000\;gr\;mm^{-1}}$). The marginalized posterior for $D$ is broader and more asymmetric than for panels 1 and 3, reflecting reduced sensitivity to duty cycle at finer groove spacing due to the limited number of propagating orders.}\label{fig:mcmc-app-panel2}
\end{figure}

\begin{figure}[H]
\centering
\includegraphics[width=0.9\textwidth]{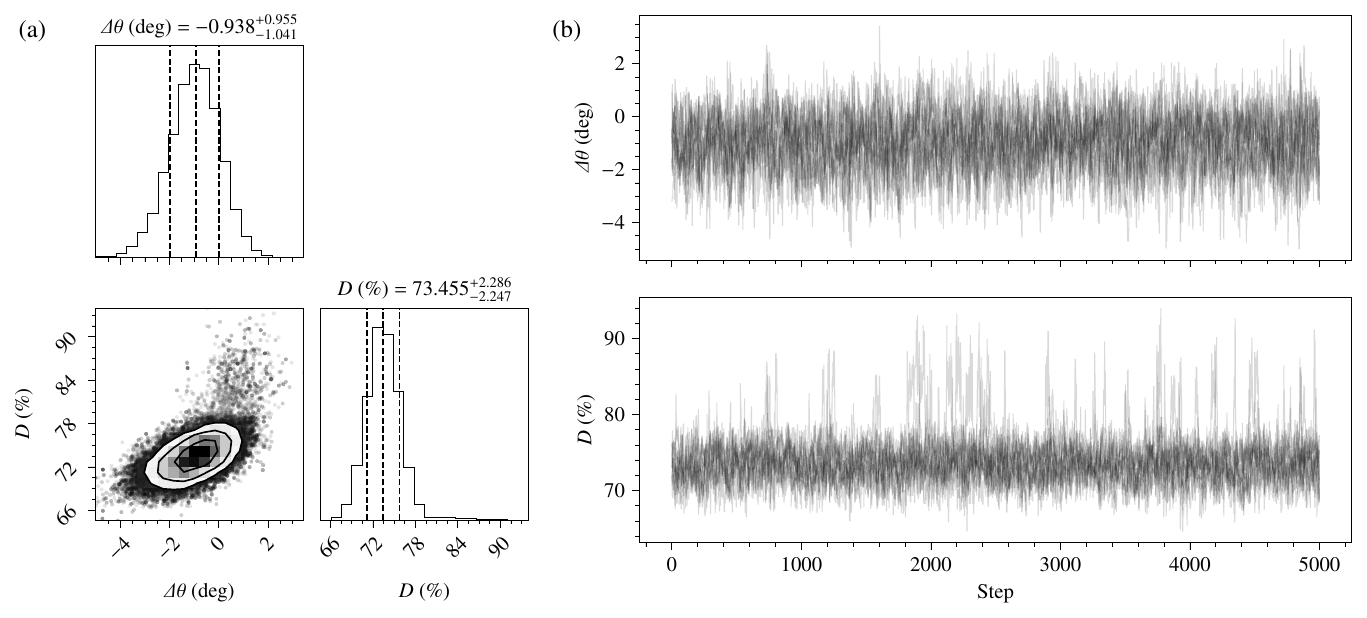}
\caption{Same as Figure~\ref{fig:mcmc-app-panel1} but for panel 3 ($\mathrm{800\;gr\;mm^{-1}}$). Retrieved parameters are consistent with panel 1, confirming that both outer panels share similar effective groove profiles, as expected for their nominal specifications and symmetric placement on the substrate relative to the central $\mathrm{2000\;gr\;mm^{-1}}$ panel.}\label{fig:mcmc-app-panel3}
\end{figure}

\end{document}